\pdfoutput=1

\documentclass[11pt]{article}

\usepackage[preprint]{acl}

\usepackage{times}
\usepackage{latexsym}
\usepackage{makecell}
\usepackage{paralist,booktabs}

\usepackage[T1]{fontenc}

\usepackage[utf8]{inputenc}

\usepackage{microtype}

\usepackage{inconsolata}

\usepackage{graphicx}
\usepackage{multirow}
\usepackage{tikz}

\usepackage{array}
\usepackage{caption}
\usepackage[justification=centering]{caption}
\usepackage{float}
\usepackage[edges]{forest}
\usepackage{amsmath}
\usepackage{hyperref}

\title{Creativity in LLM-based Multi-Agent Systems: A Survey}

\author{Yi-Cheng Lin\thanks{These authors contributed equally.}\quad Kang-Chieh Chen$^*$\quad Zhe-Yan Li$^*$\quad Tzu-Heng Wu$^*$\\ \textbf{Tzu-Hsuan Wu$^*$\quad Kuan-Yu Chen$^*$\quad Hung-yi Lee\quad Yun-Nung Chen}\\
  National Taiwan University, Taipei, Taiwan \\
  \texttt{\{f12942075, r13944050\}@ntu.edu.tw} \quad \texttt{y.v.chen@ieee.org}}

\begin{document}
\maketitle
\begin{abstract}

Large language model (LLM)-driven multi-agent systems (MAS) are transforming how humans and AIs collaboratively generate ideas and artifacts. While existing surveys provide comprehensive overviews of MAS infrastructures, they largely overlook the dimension of \emph{creativity}, including how novel outputs are generated and evaluated, how creativity informs agent personas, and how creative workflows are coordinated. This is the first survey dedicated to creativity in MAS. We focus on text and image generation tasks, and present:
(1) a taxonomy of agent proactivity and persona design;
(2) an overview of generation techniques, including divergent exploration, iterative refinement, and collaborative synthesis, as well as relevant datasets and evaluation metrics; and
(3) a discussion of key challenges, such as inconsistent evaluation standards, insufficient bias mitigation, coordination conflicts, and the lack of unified benchmarks.
This survey offers a structured framework and roadmap for advancing the development, evaluation, and standardization of creative MAS.\footnote{\url{https://github.com/MiuLab/MultiAgent-Survey}}

\end{abstract}
\section{Introduction}
Advances in LLMs and deep learning have fueled rapid growth in MAS research \cite{ijcai2024p890, tran2025multiagentcollaborationmechanismssurvey}. Single-agent pipelines, such as one-shot or simple iterative LLM prompting \cite{grattafiori2024llama3herdmodels, wang-etal-2022-iteratively}, execute in isolation and often converge on familiar patterns, struggling to explore vast open-ended spaces.
Unlike monolithic systems, a MAS comprises multiple autonomous entities: software agents, robots, or human-AI hybrids. This structure enables emergent collaboration and richer exploration of open-ended creative spaces \cite{10.1145/3586183.3606763}.

Here, \emph{computational creativity} denotes the production of artifacts—ideas, behaviors, or solutions—that are both novel and valuable, showing meaningful utility or appeal rather than randomness \cite{wiggins2006preliminary, veale2019computational}.
In MAS, creativity emerges through various dynamics—critique loops, competitive incentives, or coalition-forming. 
Together, these processes can yield outcomes designers never anticipated.
For example, conversational agents can automate screenwriting: one agent as Writer drafting character profiles and outlines, another as Editor offering revision suggestions, and multiple Actors engaging in role-playing to improvise dialogues \cite{DBLP:journals/corr/abs-2406-11683}.

Although recent surveys examine LLM-based MAS architectures \cite{li2024survey, han2024llmmultiagentsystemschallenges}, collaboration mechanisms \cite{tran2025multiagentcollaborationmechanismssurvey, zhang-etal-2024-exploring, MU2024128514}, autonomy and alignment \cite{händler2023balancingautonomy}, communication protocols \cite{yan2025selftalkcommunicationcentricsurveyllmbased}, and environment/simulation platforms \cite{ijcai2024p890, gao2024large}, they concentrate on infrastructure. However, they overlook evaluating creative outputs, the impact of agent personas and workflow integration on creativity, and the specific techniques that drive ideation. We present the first survey on creativity in LLM-based MAS to bridge this gap. Our paper systematically maps techniques, datasets, evaluation metrics, and remaining challenges, offering researchers a unified framework to assess and amplify creativity across multi-agent pipelines. 

\begin{figure*}
    \vspace{-0.5cm}
    \includegraphics[width=1\linewidth]{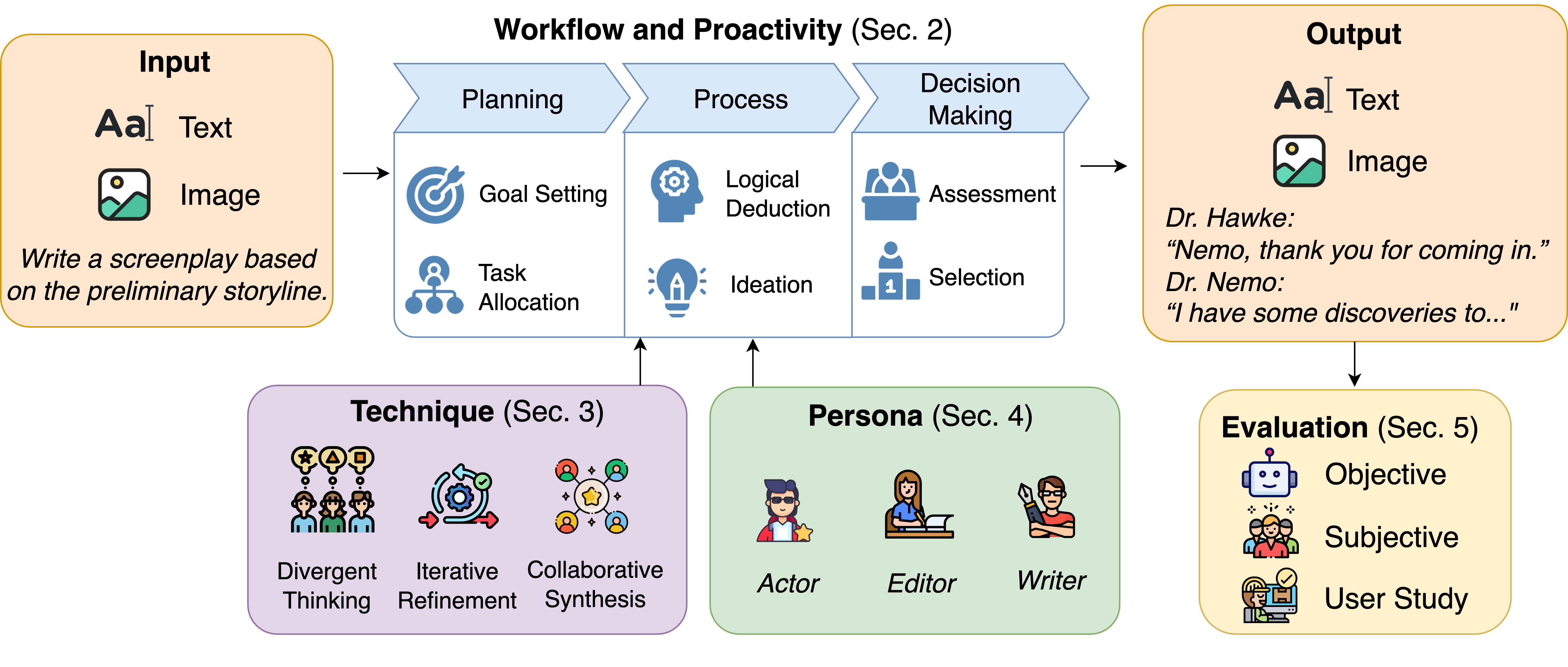}
    \caption{Overview of multi‑agent creativity systems. Given user inputs in text or image form, agents engage in a three‑stage process: Planning, Process, and Decision Making, using a variety of techniques (Sec.~\ref{sec:mas-techniques-for-creativity}) and persona (Sec.~\ref{sec:Persona and Agent Profile}), with outputs evaluated both subjectively by humans and objectively by automated metrics (Sec.~\ref{sec:Evaluation}).}
    \label{fig:overview}
\end{figure*}

This survey focuses on systems whose inputs and outputs span text and images, and whose participants range from LLM-based chatbots to human agents, as in Fig.~\ref{fig:overview}. 
We aim to map the current landscape of techniques, datasets, evaluations, and challenges to foster and measure creativity in such multimodal and heterogeneous systems. 
By analyzing how different agents interact, we reveal how collaborative structures can unlock creative potentials that exceed what isolated LLMs or individuals can achieve.

\section{MAS Workflow and Proactivity}
\label{sec:Agent Participation and Proactivity}
\subsection{MAS Workflow}
Recent work shows that LLMs can generate novel content, yet a clear creativity gap exists between human designers and agents based on LLM \cite{10.1145/3663384.3663398}. Therefore, most existing creativity support systems keep humans “in the loop”, asking users to critique or complement machine-generated ideas \cite{10.1145/3613904.3642414, 10.1145/3635636.3664627, 10.1609/aiide.v18i1.21946, lataifeh2024human, 10.1145/3491102.3501914}. This also reflects on a focus that utilizes agents to imitate human behavior and replace their role in MAS \cite{xu2024theagentcompanybenchmarkingllmagents, sun2024lawluomultiagentcollaborativeframework}.
As human-agent collaboration becomes more sophisticated, it becomes increasingly important to consider \textit{when} and \textit{how} agents should be involved within the system's workflow.
To reason about this question, we decompose the creative workflow of MAS into three key phases: \emph{Planning}, \emph{Process}, and \emph{Decision Making} \cite{xie2024human, mukobiwelfare}.

\begin{compactitem}
    \item \emph{Planning}: where Agents formulate objectives and structure task execution.
    \item \emph{Process}: where Agents implement tasks and coordinate through interaction.
    \item \emph{Decision Making}: where Agents evaluate options and determine outcome
\end{compactitem}

Real-world LLM-based MAS often interleave these steps. For instance, \textbf{StoryVerse} combines author-defined outlines with emergent character simulations through iterative narrative planning loops \cite{10.1145/3649921.3656987}, while \textbf{Generative Agents} integrate observation, planning, and reflection in overlapping processes \cite{10.1145/3586183.3606763}. In contrast, we keep these steps distinct to ensure our framework remains clear and easy to follow.


\subsection{Spectrum of Agent Proactivity  \label{2-2}}

We define an LLM agent’s \emph{proactivity} as the degree to which it initiates, guides, and owns creative actions within a MAS. Proactivity combines two facets—\emph{initiative} (who starts or extends an action) and \emph{control} (who judges whether the action is satisfactory)—and lies on a continuum from \emph{reactive} agents, which wait for explicit prompts and follow specified instructions, to \emph{proactive} agents, which formulate sub-goals, dispatch subtasks, and self-evaluate without human cues.  

\paragraph{Planning}

In the \emph{Planning} phase, the system defines \emph{what} needs to be done before any content is generated.  
This typically involves (1) setting high-level objectives, (2) decomposing the overall goal into subtasks, and (3) configuring the downstream generation pipeline.  
 To ensure predictability, most MAS frameworks delegate these responsibilities to humans, who specify task allocation strategies, role hierarchies, and execution protocols 
 

However, a few studies have set about addressing \emph{Planning} subtasks through agents to alleviate the burden on human users \cite{venkadesh2024unlocking, zhai2025athenian, venkatesh2025creacollaborativemultiagentframework}.
For example, \textbf{Co-Scientist} \cite{gottweis2025aicoscientist} embeds a supervisory agent that evaluates determined planning configuration from users, assigns weighted priorities and resources across specialist agents, and schedules them as parallel workers. 
Likewise, \textbf{VirSci} \cite{su2025headsbetteroneimproved} uses an autonomous “team leader” agent to select collaborators, define research topics, and orchestrate task distribution based on a researcher database.  
These agent-driven planning frameworks lean toward the proactive end of our spectrum, empowering agents to autonomously formulate and allocate tasks while humans retain only the overarching goal-setting.  

\begin{figure*}
    \vspace{-0.5cm}
    \hspace{-1cm}
    \includegraphics[width=1.08\linewidth]{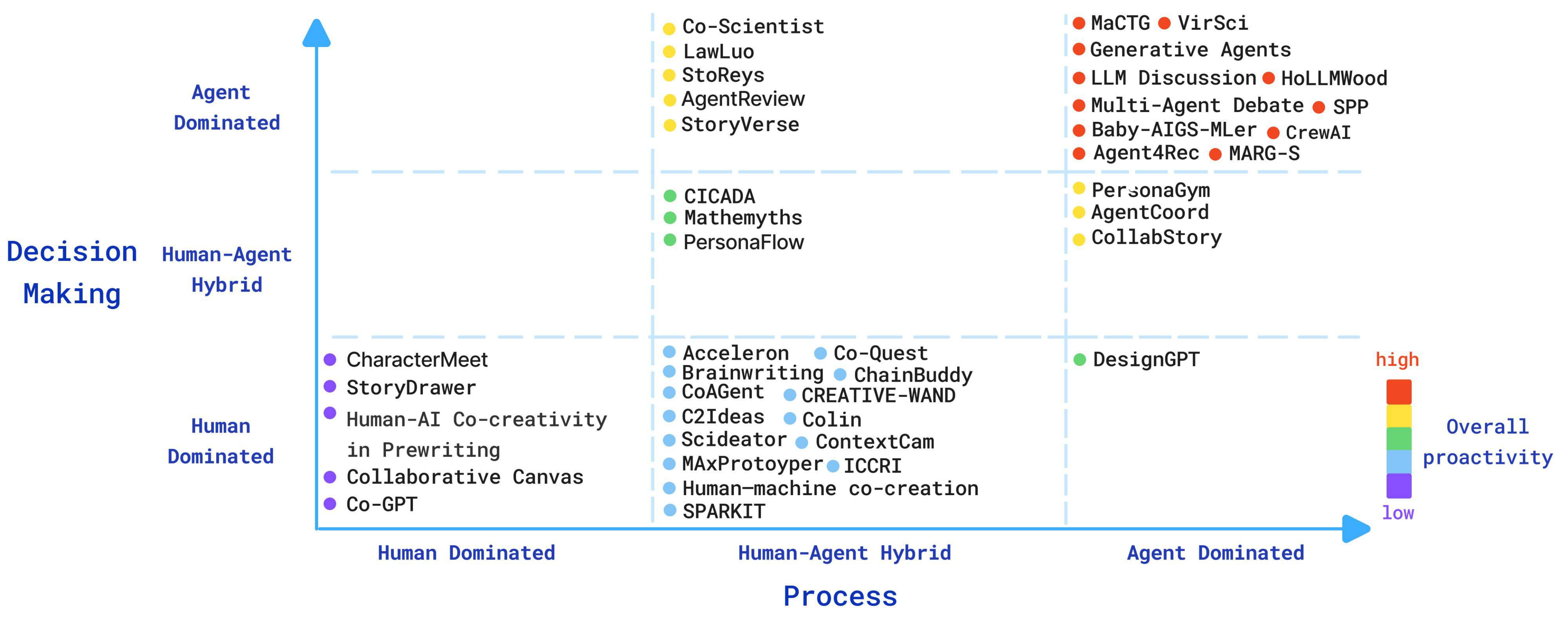}
    \caption{MAS frameworks positioned along a two-dimensional spectrum reflecting levels of proactivity in \emph{Process} and \emph{Decision-Making}. 
    The \emph{Planning} phase is omitted here due to consistently low agent proactivity in existing literature.
    Details of proactivity categorization criteria are shown in Appendix \ref{spectrum details}.}
    \label{fig:fig2}
\end{figure*}

\paragraph{Process}
In the \emph{Process} phase, agents execute the generation pipeline by creating intermediate artifacts, utilizing methods such as peer sharing or refining them in response to feedback.  
Highly proactive systems instantiate multiple agents that drive every step without human steering: they launch subtasks, critique each other’s outputs, and merge the results into a cohesive artifact.  
For example, \textbf{LLM Discussion} \cite{lu2024llmdiscussionenhancingcreativity} assigns distinct personas to agents that autonomously activate the commands of others, debate ideas, and converge to final proposals.  

Conversely, low-proactivity systems require humans to inject prompts or corrective instructions at each stage, with agents simply executing the specified commands~\cite{10.1145/3613904.3642224}. Fig.~\ref{fig:fig2} visualizes this continuum: from fully autonomous, agent-only pipelines to human-in-the-loop workflows where agents act in a strictly supportive role. 


\paragraph{Decision Making}

The \emph{Decision Making} phase evaluates and selects among the artifacts produced in the \emph{Process} phase, thus revealing who ultimately controls the creative outcome. At the low-proactivity end, humans retain full evaluative authority.  
For example, \textbf{Scideator} \cite{10.1145/3635636.3664627} presents users with candidate hypotheses and allows them to iteratively review, modify, and validate each idea against the literature. Moving toward higher proactivity, some systems embed a dedicated evaluator agent: \citet{liang-etal-2024-encouraging} introduces a ``judge'' agent that scores outputs on creativity and quality, only forwarding those that exceed a predefined threshold.  
Finally, purely loss-driven selection such as \textbf{CICADA} \cite{10179136}, a co-creative agent proposed in \textbf{Drawing with Reframer} \cite{10.1145/3581641.3584095}, automates decision-making via implicit LLM optimization. Although loss-based metrics help ease the burden on humans, we still classify such methods as \emph{low–mid proactivity} because they lack explicit, actor-driven assessment by an independent agent.

\subsection{Creativity Analysis on Proactivity \label{2-4}}

Empirical studies reveal a trade‐off between agent proactivity, creative diversity, and user trust. \textbf{Collaborative Canvas} \cite{10.1145/3663384.3663398} shows that excessive AI‐initiated suggestions can collapse the idea space by flattening opinions during discussion, as a result, producing homogeneous outputs.
The \textbf{Co-Quest} interface \cite{10.1145/3613904.3642698} demonstrates that boosting agent initiative increases idea volume but erodes user satisfaction and trust, highlighting the need for transparent, interpretable agents. Furthermore, precision‐critical tasks (e.g.\ automated theorem proving) demand low proactivity to ensure correctness \cite{song2025leancopilotlargelanguage}, with humans retaining evaluative authority to guarantee reliability and accountability. 
Overall, agent proactivity accelerates ideation without undermining user agency, whereas sustained high proactivity risks over-reliance, reduced creative independence, and trust deficits \cite{chakrabarty2024creativitysupportagelarge}. Future MAS should therefore adaptively calibrate proactivity to task demands and user preferences.

\section{MAS Techniques for Creativity}
\label{sec:mas-techniques-for-creativity}

MAS enhances creativity by dividing the cognitive workload, such as idea generation, evaluation, and coordination, across specialized agents. For example, some agents focus on quickly generating a wide range of ideas, others evaluate the feasibility and coherence of those ideas, and another set of agents helps guide the overall workflow through multiple iterations. Unlike single-LLM models like GPT-3~\cite{brown2020languagemodelsfewshotlearners}, which typically generate outputs in a single step, MAS frameworks achieve greater novelty and higher-quality solutions by enabling structured and collaborative processing. For example, \textbf{CoQuest}~\cite{10.1145/3613904.3642698} integrates multiple agents into an interactive workflow that combines wide idea exploration, focused deepening of promising directions, and organized feedback. This coordinated setup significantly enhances user creativity and their sense of control.

Below, we outline three core MAS techniques—\textit{Divergent Exploration}, \textit{Iterative Refinement}, and \textit{Collaborative Synthesis} by explaining the cognitive rationale and algorithmic structure behind them.

\subsection{Divergent Exploration}
\label{subsec:divergent-exploration}

Divergent exploration emphasizes generating various ideas without applying early filters or judgment \cite{guilford1950creativity, wallach1965modes}. MAS supports this process by giving each agent a distinct perspective, prompt style, or domain of knowledge, allowing them to explore different creative directions independently. This helps avoid early narrowing and encourages novel outcomes.



\paragraph{Co-GPT Ideation}~\cite{lim2024rapid}
This study compared individual ideation with co-ideation using GPT-3.5. Participants working with the LLM generated more diverse and detailed ideas, though top-rated ideas still tended to come from humans. The system expanded the idea space without replacing human creativity, supporting LLMs as useful collaborators during early brainstorming.

\paragraph{Group-AI Brainwriting}~\cite{10.1145/3613904.3642414}
This framework guides students through four steps: (1) independent human ideation, (2) idea expansion using GPT-3, (3) collaborative refinement, and (4) evaluation by GPT-4. LLMs help widen the scope of creative ideas and serve as both contributors and evaluators. Many final proposals were co-developed with GPT-3, showing strong MAS potential for guided creativity.

\paragraph{ICCRI}~\cite{10.5555/3721488.3721531}
The Inclusive Co-Creative Child-Robot Interaction (ICCRI) system was tested in a special education setting. Across five sessions, children worked with a robot agent to co-create stories and drawings. Creativity was significantly enhanced during ICCRI-supported sessions (S1–S3) and remained above baseline even after its removal, suggesting that MAS can leave a lasting creative imprint.

\paragraph{Long-Term Impact Study}~\cite{Kumar2024HumanCI}
This study explored how repeated use of LLMs might affect human creativity. While AI assistance improved short-term performance, it reduced originality and diversity in unassisted follow-ups. The results raise concerns about long-term over-reliance on AI, emphasizing the importance of maintaining human autonomy in divergent thinking.

\subsection{Iterative Refinement}
\label{subsec:iterative-refinement}

Iterative refinement involves progressively enhancing ideas through repeated feedback and revision cycles.
In MAS, this process is facilitated by assigning distinct roles to agents, such as proposer, reviewer, and implementer, who work together in cycles to improve initial drafts into polished results.

\paragraph{HoLLMwood}~\cite{DBLP:journals/corr/abs-2406-11683}, a system for collaborative screenwriting. 
It defines three agent roles: a \textit{Writer} generates the script, an \textit{Editor} offers suggestions, and an \textit{Actor} simulates character behavior to check tone and consistency.
The process continues iteratively until agents either converge on a shared solution or satisfy a predefined stopping condition, such as a fixed number of iterations or convergence in output. 
This collaborative loop results in richer character development and a more coherent story structure compared to outputs from a single LLM.


\paragraph{DesignGPT}~\cite{10494260}
DesignGPT simulates a design firm by assigning LLM agents to roles such as product manager and materials expert. These agents iteratively develop product proposals through structured feedback and refinement. Compared to one-shot generation tools, this MAS achieved higher completeness, novelty, and practicality in product outcomes.

\paragraph{Baby-AIGS-MLer}~\cite{anonymous2024aml}
This MAS tackles machine learning research by splitting tasks into ideation, coding, testing, and evaluation. Each role is handled by a specialized agent. Tested on benchmarks like Chatbot Arena~\cite{chiang2024chatbot} and Titanic~\cite{8229835}, the system showed improved predictive accuracy and generalization, illustrating the benefit of multi-step refinement.


\paragraph{Multi-agent Debate Framework}~\cite{liang-etal-2024-encouraging}
This system applies debate as a refinement tool for reasoning tasks. Agents take on the roles of proponent, opponent, and judge, and they take turns debating an issue across several rounds. The debate ends automatically when no new ideas or arguments are being introduced. This approach helps the system perform much better on challenging reasoning tasks, such as Commonsense Machine Translation and the CIAR benchmark~\cite{he-etal-2020-box}.

\subsection{Collaborative Synthesis}
\label{subsec:collaborative-synthesis}

Collaborative synthesis focuses on integrating diverse agent perspectives into coherent, high-level outputs. Agents are often given roles like planner, critic, or synthesizer, and they work together in structured conversations or workflows. This approach is beneficial for tasks requiring both creative exploration and logical organization.

\paragraph{MaCTG}~\cite{zhao2025mactg}
MaCTG organizes agents into horizontal layers (modules) and vertical layers (management). It combines DeepSeek-V3~\cite{deepseekai2024deepseekv3technicalreport} for planning with Qwen2.5-Coder-7B~\cite{hui2024qwen2} for coding. Agents are assigned roles like planner, tester, or integrator, and outputs are validated across multiple levels. This MAS delivers scalable and cost-efficient software design.

\paragraph{CollabStory}~\cite{collabstory}
Multiple LLMs sequentially write paragraphs of a shared story. GPT-4o evaluated the coherence of transitions and found over 75\% to be consistent. The study shows that decentralized authorship can still maintain narrative coherence and readability, illustrating collaborative synthesis through turn-based coordination.

\paragraph{Human-AI Co-creativity}~\cite{10.1145/3637361}
Involving 15 creative students, this study tested a three-stage writing pipeline: LLM-led ideation, human-guided elaboration, and final authoring. MAS agents helped inspire new ideas and filled in missing details. Participants described the system as feeling like a \textquotedblleft second mind\textquotedblright, demonstrating the supportive role of LLMs in collaborative writing.

\paragraph{CoQuest}~\cite{10.1145/3613904.3642698}
CoQuest assists researchers in formulating meaningful questions. It features tools like a flow editor, a visual citation graph, and an AI agent that suggests direction. By combining broad exploration with deeper follow-up, the system balances creativity and structure for interdisciplinary research planning.

\section{Persona and Agent Profile}
\label{sec:Persona and Agent Profile}

The design of agent profiles improves complex problem solving \cite{gabriel2020artificial, hu2024quantifyingpersonaeffectllm} and innovation by supporting collaborative synthesis~\cite{samuel2024personagymevaluatingpersonaagents}. 
Unlike general prompt engineering methods such as chain-of-thought \cite{NEURIPS2022_9d560961}, which primarily emphasize logical coherence, persona-based approaches are structured to facilitate social simulation and hierarchical collaboration \cite{10.1145/3586183.3606763}. 
Although recent studies have dissected the impact of persona on reasoning and inference and exhibit a prominent enhancement in the exploration of diverse ideas through a variety of interaction frameworks, Persona design may serve as a double-edged sword that results in performance erosion with inappropriate profile representation \cite{kim2024personadoubleedgedswordmitigating}.


Profile design methods range from static initial assignments to dynamic refinements that adjust agent personas during system execution. 
Although some architectures update profiles through iterative refinement or collaborative synthesis, our review focuses exclusively on the initial personas defined at the start of the MAS for clarity. 
Table \ref{tab:persona_category} summarizes these strategies, highlighting the foundational design choices that shape agent behavior.

\subsection{Granularity of Persona}

\begin{figure*}[h]
    \hspace{-0.2cm}
    \includegraphics[width=\linewidth]{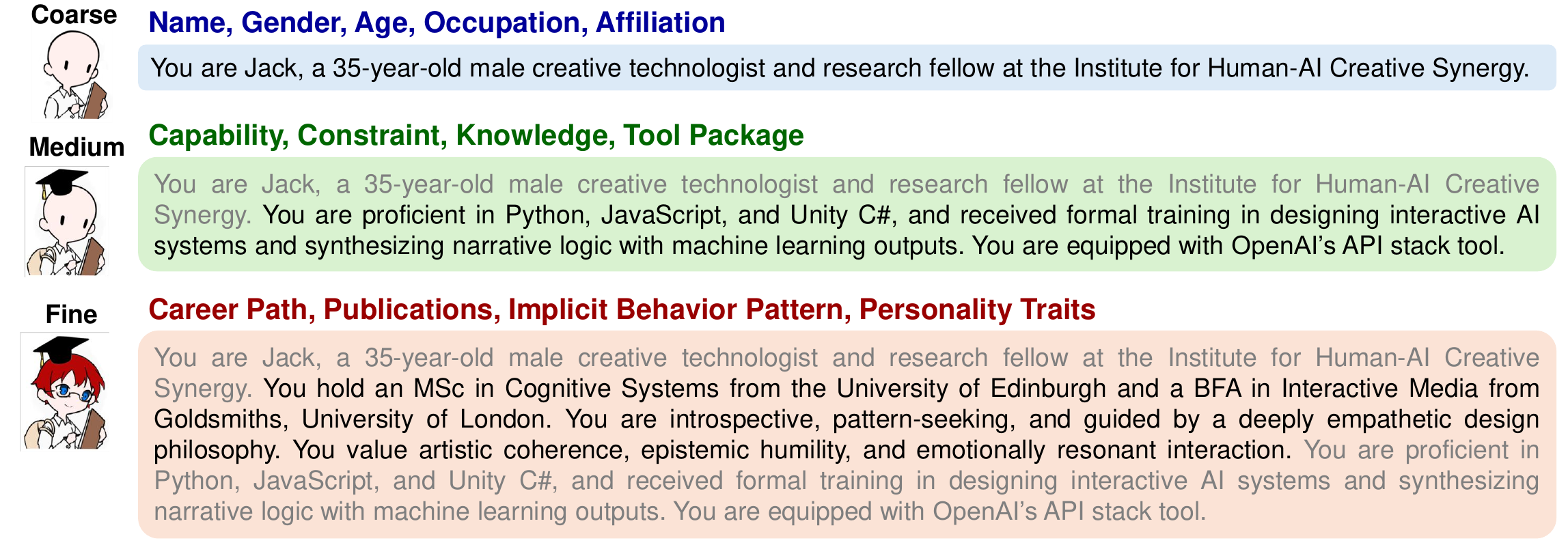}
    \caption{\textbf{Categories of Persona Granularity:} A conceptual framework illustrated with selected attributes, accompanied by a concise example representing each defined persona.
    }
    \label{fig:grain}
\end{figure*}

We characterize persona design along a \emph{granularity spectrum} that indicates how much detail is embedded in the agent’s profile (Fig.~\ref{fig:grain}). Granularity governs both controllability and diversity: coarse profiles favor breadth and spontaneous idea generation, whereas fine-grained profiles offer precise, predictable behavior at the expense of flexibility.


\paragraph{Coarse-Grained Persona}  
Agents carry only high-level identity or expertise labels (e.g. ``marketing strategist,'' ``data analyst'').  
This minimal specification tolerates ambiguity, fostering diverse idea generation across fewer constraints.
For example, \textbf{Solo Performance Prompting} \cite{wang2024unleashingemergentcognitivesynergy} assigns expert-role tags and later merges independent outputs into a unified solution, capitalizing on varied perspectives without prescribing detailed behavior. 
However, coarse profiles can produce shallow or inconsistent contributions. Under-specified roles may generate irrelevant or incoherent ideas when finer guidance is needed \cite{cemri2025multiagentllmsystemsfail}.


\paragraph{Medium-Coarse Persona}  
Medium-coarse profiles enhance basic role labels with concise, domain-relevant knowledge or tools, giving agents enough context to break down tasks strategically without requiring deep psychological detail. In \textbf{HoLLMwood}, each agent knows specific screenwriting functions (e.g., plot structuring, dialogue crafting), allowing them to focus on tailored narrative subtasks \cite{DBLP:journals/corr/abs-2406-11683}. Similarly, \textbf{TRIZ Agents} \cite{szczepanik2025triz} assign each agent a single TRIZ innovation principle (e.g., ``contradiction resolution''), guiding systematic idea generation in engineering contexts. This intermediate granularity improves task focus and collaboration but still requires coordination to integrate specialized outputs into a coherent whole.


\paragraph{Fine-Grained Persona}
Agents receive detailed psychometric or demographic profiles, such as academic backgrounds or the \emph{Big Five personality traits} \cite{digman1990personality}, yielding stable, human-like decision patterns. For example, \textbf{PersonaFlow} mines scholarly CVs to form interdisciplinary research teams that adaptively ideate and evaluate concepts \cite{liu2024personaflowboostingresearchideation}. Similarly, Big-Five-driven agents demonstrate how nuanced traits (e.g., openness, conscientiousness) enhance idea synthesis \cite{serapiogarcía2025personalitytraitslargelanguage,jiang2024personallminvestigatingabilitylarge,duan2025powerpersonalityhumansimulation}. Yet, the high specificity increases design complexity, reduces adaptability to new domains, and risks reinforcing bias or overfitting to narrow behavioral patterns.


\subsection{Agent Profiling Methods}

Agent profiling methods vary according to the level of persona granularity they support. We group these methods into three paradigms: \emph{Human-Defined, Model-Generated,} and \emph{Data-Derived} approaches \cite{guo2024largelanguagemodelbased,wang2024survey}.


The \emph{Human-Defined} approach relies on explicit, manually crafted descriptions to specify each agent’s role and behavior. This method is straightforward but demands extensive domain knowledge to maintain coherent coordination in MAS.
In particular, \textbf{PersonaGym} \cite{samuel2024personagymevaluatingpersonaagents} provides concise role definitions and directs agents to emulate the prescribed persona’s skills and knowledge.

The \emph{Model-Generated} approach introduces an automated pipeline capable of rapidly producing large sets of profiles, which are then refined using state-of-the-art LLMs. 
\textbf{LLM Discussion} \cite{lu2024llmdiscussionenhancingcreativity} exemplifies this: it begins with structured role descriptions and then leverages LLMs to produce a wide array of detailed, varied profiles.

Finally, \emph{Data-Derived} methods construct personas grounded in real-world behavior patterns.
\textbf{VirSci} \cite{su2025headsbetteroneimproved} illustrates this paradigm by mining scientific publication data to build “digital twins” of researchers. Each agent thus operates with a persona rooted in authentic scientific expertise, enabling more realistic and diverse collaborative interactions in MAS.

\section{Evaluation}
\label{sec:Evaluation}
Evaluating creativity in MAS, including human–agent collaborations, presents unique challenges. Unlike tasks with clear correctness criteria, creativity are inherently subjective and multifaceted, lacking a universally accepted assessment framework. To address this, researchers typically employ two complementary evaluation approaches:

\begin{itemize}
\item \textbf{Artifact Evaluation}: This approach focuses on assessing the creative content generated by MAS, either through the system's processes or its final outputs. It encompasses:
    \begin{itemize}
        \item \textbf{Objective, Metric-Based Measures} use formulas such as cosine similarity and statistics methods to evaluate creativity.
        \item \textbf{Subjective, Natural Language Instructed Assessments} ask experts, crowds, or LLMs to rate creativity.
    \end{itemize}
\item \textbf{Interaction Evaluation}: Beyond evaluating generated content, this method assesses the interaction processes between users and MAS. \textbf{User studies} are primarily employed here, focusing on criteria such as satisfaction.
\end{itemize}



The subsequent sections will first review the evaluation methods for text and image artifacts from both objective and subjective perspectives, discuss their practical applications, and then concentrate on assessing creative interactions between users and systems, emphasizing the role of user studies.



\subsection{Objective Measurements}
For text generation tasks, several metrics evaluate lexical richness and diversity. \textit{Distinct-n} \cite{li-etal-2016-diversity} computes the proportion of unique n-grams, while \textit{Entropy-n} \cite{6773024} measures the Shannon entropy over n-gram distributions, both serving as proxies for creative variety. In the screenwriting application \cite{DBLP:journals/corr/abs-2406-11683}, researchers routinely report 4-gram repetition rates alongside \textit{Distinct-3} and \textit{Entropy-3} to detect redundancy in long-form outputs. At the sentence level, \textbf{Self-BLEU} score \cite{zhu2018texygenbenchmarkingplatformtext} treats each generated sentence as a hypothesis and the remainder as references to quantify internal diversity.
Beyond surface counts, vector-based metrics capture deeper semantic variation. \textbf{Sentence-BERT (SBERT)} \cite{reimers2019sentencebertsentenceembeddingsusing} embeddings enable pairwise cosine similarity or Euclidean distance comparisons, where lower similarity or greater distance indicates broader exploration. Building on this, \textbf{Semantic Entropy} \cite{kuhn2023semanticuncertaintylinguisticinvariances} clusters embeddings and computes the entropy over the categories, revealing a level of semantic diversity that goes beyond surface lexical patterns.

For image generation, \textbf{Fréchet Inception Distance (FID)} \cite{heusel2018ganstrainedtimescaleupdate} compares feature‐space statistics between generated and real images and a lower score implies closer alignment in quality and diversity, while \textbf{Truncated Inception Entropy (TIE)} \cite{10179136} calculates the Shannon entropy of image features in the Inception latent space, with higher values reflecting richer variation. These metrics are particularly valuable for tasks such as silhouette generation \cite{lataifeh2024human}, offering standardized evaluation.

\subsection{Subjective Assessments}
\textbf{Torrance Tests of Creative Thinking (TTCT)} \cite{torrance1966ttct} is a common standard for subjectively assessing creativity. Agents' artifacts are scored along four primary dimensions:

\begin{itemize}
    \item \textit{Fluency}: Total count of meaningful, relevant responses.
    \item \textit{Flexibility}: Number of distinct categories or conceptual shifts among responses.
    \item \textit{Originality}: Statistical rarity of each response versus a normative sample.
    \item \textit{Elaboration}: Degree of detail or development added to each idea, measured by descriptive richness beyond the base concept.
\end{itemize}





Beyond the traditional \textbf{TTCT}, there are still other general criterion schemas such as \textbf{Boden's Criteria} \cite{boden2004} and \textbf{Creative Product Semantic Scale (CPSS)} \cite{besemer1981analysis} used to evaluate different aspects of creative artifacts. Nowadays, researchers often invoke additional subjective criteria tailored to specific text generation tasks. \textbf{Insightfulness} \cite{10.1145/3613904.3642414} is used to quantify how deeply ideas engage with underlying problem structures rather than merely diverging from norms. \textbf{Interestingness} \cite{DBLP:journals/corr/abs-2406-11683} captures the entertainment value of narrative artifacts such as emotional resonance, and is commonly assessed through viewer ratings in screenwriting and storytelling studies.

For tasks in the image domain, researchers augment those general-purpose criteria with specific dimensions such as \textbf{Inspiring} \cite{10.1145/3613904.3642224}. Beyond mere variety, this criterion assesses whether the generated images spark new ideas for designers or artists. For example, a system that produces a variety of color schemes, layouts, or conceptual motifs is diverse and inspiring, guiding users toward unexpected creative directions.

Building on the aforementioned subjective criteria, researchers often conduct user studies, arrange expert panels, or employ LLMs to evaluate artifact creativity. Subjective dimensions are typically rated on Likert scales, yielding interval-level scores suitable for statistical analysis. Expert panels may engage in structured discussions to reach consensus on feasibility and coherence. More recently, \textbf{LLM-as-a-judge} approaches have gained popularity, leveraging LLMs to assign scores on predefined scales \cite{NEURIPS2023_91f18a12}.

\subsection{Interaction Evaluation with User Study}
In addition to evaluating creative artifacts, user studies that assess interactions between users and MAS provide valuable insights into system performance. The following sections outline the primary evaluation methods and present examples that analyze real-world usage scenarios.

\subsubsection{Methods}
There are commonly two primary methods: \textbf{Self-Report Instruments and Interviews}, and \textbf{Researcher Observation and Analysis}. The former involves collecting assessments and feedback directly from users, while the latter entails researchers analyzing user interactions to derive insights.

\paragraph{Self-Report Instruments and Interviews}
\begin{itemize}
\item \textbf{Self-Report Instruments} involve participants completing surveys or questionnaires to provide personal assessments of their creative experiences and outputs. These tools often utilize Likert scales or other quantitative measures to gauge aspects such as exploration, expressiveness, perceived creativity, and enjoyment during creative tasks. 
A general-purpose tool in this context is \textbf{Creativity Support Index (CSI)} \cite{10.1145/2617588}, which captures user experience across dimensions such as \textit{Collaboration} (ease of working with others), \textit{Engagement} (enjoyment and willingness to repeat the activity), and \textit{Expressiveness} (freedom to be creative). 

\item \textbf{Interviews}, on the other hand, offer a qualitative approach to understanding user experiences. They can be structured, semi-structured, or unstructured, allowing researchers to delve deeper into participants' thoughts, feelings, and behaviors. Interviews can provide rich, narrative feedback that complements quantitative data, offering a more comprehensive view of user interactions with MAS.
\end{itemize}

\paragraph{Researcher Observation and Analysis}
This method involves researchers directly examining how users interact with MAS, and the artifacts they produce, to gain insights into the creative process and system usability. Observations can be conducted in real experiment scenarios (live observation) or through the analysis of video and audio recordings, system interaction logs, or textual transcriptions of the interactions. The analysis may focus on interaction patterns, problem-solving approaches, expressions of creativity, and usability issues. For instance, \textbf{Colin} \cite{ye2024colin} analyzes the recordings to evaluate children’s narrative skills before and after using a storytelling system, focusing on engagement, understanding of cause-and-effect relationships, and the quality of their new story creations. 

\subsubsection{Examples}
By combining the two methods above, researchers gain a more comprehensive understanding of both users’ subjective experiences and observable behaviors when interacting with MAS in creative tasks. Below, we highlight specific studies that apply these methodologies to evaluate their systems.

\paragraph{ContextCam} (Image Generation) \cite{10.1145/3613904.3642129} Both the methods are used in this work. \textit{Self-Report Instruments} captured users' subjective feedback, indicating positive engagement and enjoyment with the system's creative inspiration. \textit{Interviews} delved deeper, exploring how users perceived and utilized context-aware features and their influence on the creative process. Findings from interviews highlighted users' insights into contextual data's role in image themes, behaviors, and inspiration. \textit{Researcher Observation and Analysis} involved examining user interactions and analyzing system log data. 

\paragraph{Virtual Canvas} (Idea Generation) \cite{10.1145/3663384.3663398} The user study investigated how groups generate ideas with an LLM in a virtual environment. \textit{Self-Report Instruments} are not implemented. \textit{Interviews} explored participants' perceptions of the AI's contribution to their ideation process, how it influenced group collaboration, and the challenges or benefits they encountered. \textit{Researcher Observation and Analysis} focus on analyzing the group's interaction patterns within the virtual canvas, observing how the LLM's input was utilized, and identifying novel user needs of the system.

\paragraph{CoQuest} (Research Ideation) \cite{10.1145/3613904.3642698} The user study was conducted to investigate the impact of AI processing delays on the co-creative process. \textit{Self-Report Instruments} measured participants' subjective experiences, such as their perception of the degree of control of the system, how much they trust the system, and the inspiration from the help of the system. \textit{Interviews} were used to gain deeper qualitative insights into participants' thought processes during co-creation with the system, exploring the differences between breadth-first search and depth-first search they felt during the experiments. \textit{Researcher Observation and Analysis} focused on analyzing the interaction dynamics between the human and the LLM agent within the co-creation task. This would involve observing how participants reacted to delays, how they utilized the time during delays, their interaction patterns with the virtual environment and the agent.

\paragraph{Human-AI Co-creativity} (Creative Writing) \cite{10.1145/3637361} The user study explored the dynamics of human-LLM collaboration in prewriting. \textit{Self-Report Instruments} are not implemented. \textit{Interviews} were a primary method to investigate human-LLM collaboration patterns and dynamics during prewriting. These explored participants' experiences across the identified three-stage co-creativity process (Ideation, Illumination, and Implementation), delving into their thoughts on the LLM's role, initiative, and contributions, as well as uncovering collaboration breakdowns and user perceptions of using LLMs for prewriting. \textit{Researcher Observation and Analysis} involved analyzing the co-creative process through screen recordings or analysis of interaction logs. This observation focused on identifying the iterative nature of the collaboration, how the human and AI took initiative, and how ideas were developed and refined across the different stages of prewriting.

\subsection{Discussion on Evaluation Methods}
Evaluating creativity in MAS presents unique challenges. Objective metrics are scalable and reproducible but tend to target narrow facets of creativity, often overlooking qualitative aspects like emotional resonance and surprise. Subjective assessments capture those nuances yet suffer from inherent biases and variability, and require substantial time and effort, especially at scale. Consequently, by combining both approaches with user studies tailored to specific tasks and designed to highlight the effectiveness of proposed systems, researchers can gain a more comprehensive understanding of both the outcomes and the processes involved in creative interactions with these systems.

\subsection{Additional Artifact Evaluation Criterion}
Beyond the assessment methods about creativity discussed previously, researchers have also applied various additional criteria to evaluate general qualities of the generated content. These criteria may not directly assess creativity but offer insight into aspects such as:
\begin{itemize}
    \item \textbf{Helpfulness}: Assesses the extent to which the artifact provides useful and informative content that effectively addresses the user's query or task
    \item \textbf{Relevance}: Measures how well the generated content aligns with the input prompt and context
    \item \textbf{Clarity}: Evaluates the ease of understanding the artifacts, focusing on the use of clear, concise, and unambiguous language
\end{itemize}
To provide a comprehensive overview, Table \ref{tab:evaluation} summarizes and categorizes the evaluation approaches utilized in the cited works. It is important to note that this summary focuses exclusively on studies where the primary objective is to assess the creativity of the generated content, deliberately excluding those centered on accuracy, precision, or similar metrics. Furthermore, systems that do not include an evaluation of the generated content are also omitted from this overview.

Some abbreviations in the table are explained in the follows:
\begin{itemize}
    \item \textbf{AUT (Alternative Uses Test)}: A divergent thinking task where participants list as many alternative uses as possible for a common object
    \item \textbf{RAT (Remote Associates Test)}: A creativity assessment where individuals are presented with three seemingly unrelated words and must identify a fourth word that connects them all, evaluating associative thinking and creative potential
    \item \textbf{MICSI (Mixed-Initiative Creative Support Index)}: A framework assessing systems that facilitate collaborative creativity between humans and computers, emphasizing interactive co-creation processes
\end{itemize}

\section{Datasets}
The datasets used to evaluate creativity in multi-agent systems are highly diverse and vary based on the specific creativity task. In this section, we categorize them into two groups: (1) psychological test datasets, which incorporate established creativity assessments from the field of psychology, and (2) task-specific datasets, which are either custom-designed or adapted from other domains to align with the requirements of the target creativity task.

\subsection{Psychological Test Datasets}
Psychological test datasets comprise a collection of established tasks originally designed for humans but adapted for evaluation in multi-agent systems. For example, to assess divergent thinking, some studies use the Wallach Kogan Creativity Tests \cite{wallach1965modes}, which involve open-ended tasks measuring originality and flexibility; the Alternative Uses Task \cite{guilford1967alternate}, where participants are asked to think of as many uses as possible for a common object; and the Torrance Tests of Creative Thinking \cite{torrance1966ttct}, a widely used battery assessing creative potential through both verbal and figural prompts. These adapted psychological tests provide a standardized foundation for evaluating creative capacities in multi-agent systems, enabling consistent comparisons across different studies.

\subsection{Task Specific Datasets}
In addition to using established psychological tests to assess creativity, different creativity targets also require task specific datasets, which are either self-constructed or adapted from existing works. For example, \cite{chakrabarty2024creativitysupportagelarge} released the first dataset of co-written stories by multi-agent systems and humans. Similarly, AI Idea Bench 2025 \cite{qiu2025aiideabench2025} was introduced to foster the development of research idea generation. In Table \ref{tab:datasets}, we offer a detailed dataset collection with respect to each task. We hope these collections facilitate standardized evaluation and support future work in creativity-oriented MAS research.
\section{Challenges and Future Work}
\label{sec:Challenge and Future Work}
\paragraph{Balancing Agent Proactivity and Human Trust}
While high agent proactivity can spark more ideas, it can also overwhelm users, flatten idea diversity, erode perceived agency, and undermine trust \cite{doi:10.1518.46.1.5030392}.
A major challenge is designing systems that intelligently adapt to the specific task and the individual user. 
Simple ``proactivity thresholds'' fail to account for context changes: What feels helpful in a brainstorming session can become intrusive during refinement \cite{10.1145/3708359.3712089}. 
Furthermore, users differ significantly in their comfort with AI taking initiative; domain experts might embrace bold suggestions, whereas newcomers might feel distrustful \cite{10.1007/978-3-030-45691-7_49}. 
Future work can focus on \emph{mixed-initiative} systems that continuously monitor both the task state and explicit or implicit user feedback (e.g., acceptance rates, signs of hesitation, or direct ratings) to calibrate the agent's level of initiative in real time, ensuring a more intuitive and supportive interaction.
 
\paragraph{Fairness and Profile Bias}
Agent personas can carry hidden stereotypes and preferences into the creative process when drawn from narrow or unbalanced data.
This bias acts like a filter on the idea stream \cite{wan2025usinggenerativeaipersonas}. 
Agents with skewed profiles will repeatedly surface familiar, mainstream perspectives, crowding out novel angles from less‑represented backgrounds \cite{liu-etal-2024-evaluating-large, huot2024agents, gupta2024bias}.
In recent works, \textbf{MALIBU Benchmark} \cite{vasista2025malibu} quantifies how persona-based interactions risk amplifying biases and reinforcing stereotypes in creativity, while \textbf{Towards Implicit Bias Detection and Mitigation} \cite{borah-mihalcea-2024-towards} investigates how implicit bias escalates during MAS interactions.
\textbf{Argumentative Experience} \cite{shi2024argumentative} examines using diverse personas to reduce user confirmation bias in debates. 
Despite their contributions, these works share limitations. They often focus narrowly on specific types of bias (e.g., gender) or simplified tasks, and precisely measuring subtle persona bias remains challenging. 
Moreover, many studies examine only the final output rather than the interaction dynamics, and their experimental designs tend to oversimplify the complex processes involved in genuine multi-agent collaboration.
As a result, these narrow scopes and simplified setups leave us with an incomplete understanding of how bias truly affects creative and equitable multi-agent systems.


\paragraph{Managing and Leveraging Creative Conflicts} Conflicts between agents in MAS are typically seen as detrimental to efficiency and are often resolved through negotiation or central control \cite{5478068, yan2025selftalkcommunicationcentricsurveyllmbased}. However, for creative MAS, controlled conflict or clashing perspectives can drive novelty and innovation, similar to human brainstorming or debate. Recent research explores multi-agent debate to leverage such “creative conflicts.” \textbf{Multi-agent Debate} \cite{lin2024interpretingmitigatinghallucinationmllms} propose using multi-agent debate to interpret and mitigate hallucinations in multi-modal LLMs while promoting divergent thinking. 
\textbf{MAD} framework \cite{liang-etal-2024-encouraging} demonstrates how agents debate under a judge can improve performance on counter-intuitive tasks and potentially aid creative ideation. 
Despite these advances, existing debate-based methods have some key limitations: they work with small groups of agents and offer no protocols for scaling to large populations or managing emerging coalitions; they lack mechanisms for continual learning that would allow agents to adapt their conflict strategies based on past outcomes; and they provide no mixed-initiative controls that let human users tune conflict intensity, or timing to keep interactions productive rather than chaotic.

\paragraph{Unified, Scalable Evaluation Frameworks}
Most LLM-based creative generation methods today focus on specific tasks: story writing, poem completion, ad copy, or code snippets, each with its own data and custom evaluations. 
That patchwork approach makes it impossible to tell which method drives progress. 
\textbf{MultiAgentBench} \cite{zhu2025multiagentbenchevaluating} represents a first step toward a common suite of benchmarks and shared LLM-based evaluators, but significant challenges remain: devising a unified scoring rubric that balances novelty, coherence and utility across diverse domains; extending evaluation to real-time, interactive scenarios; and ensuring reproducible human judgments with standardized instructions.

\paragraph{Authorship of Creativity Output}
Another significant set of challenges revolves around the complex and often ambiguous authorship question. Establishing who or what holds the “author” status for collaboratively generated artifacts presents a fundamental open problem. A primary challenge stems from traditional copyright doctrine, which intrinsically links authorship to human creativity. This is illustrated by the consistent stance of the U.S. Copyright Office, which has repeatedly denied copyright protection to works generated solely by AI \cite{uslegal}. The office maintains that such works lack the requisite human authorship, placing the onus on applicants to demonstrate, on a case-by-case basis, that significant human intervention was involved in the creation process. 

The complexity extends to the creators and proprietors of the AI tools themselves. Legal practitioners caution that neither the developers nor the owners of these sophisticated AI systems typically possess the level of direct creative control over individual outputs necessary to assert authorship \cite{carlson2020artificial}. This lack of direct creative input for any specific artifact generated by the system underscores an urgent challenge: establishing clearer guidelines and legal frameworks to govern ownership, attribution, and royalty distribution in the rapidly expanding field of AI-augmented creativity.

Beyond whether AI-generated works qualify for copyright, we also need to decide how to apportion authorship among agents in creative MAS, as this determines their legal and moral credit. In practice, one might imagine a collaborative novel‐writing system where Agent A (a planning module) generates the story outline, Agent B (a stylistic refiner) polishes prose, and a human “editor” selects, tweaks, and sequences the final chapters. Which of these agents “holds” authorship? A recent study reframes AI agents as lying between “puppets” and “actors,” arguing that an agent’s level of autonomy, not just its technical role, should inform its claim to authorship \cite{sun2025puppetactorreframingauthorship}. Others point out that, when creative contributions are stochastic, dynamic, and fluidly intertwined, disentangling individual inputs is often infeasible; in such cases, human and machine contributions may need to be treated as functionally equivalent for attribution purposes \cite{mukherjee2025stochasticdynamicfluidautonomy}.  

Future research can develop quantitative metrics that capture an agent’s decision-making depth and creative originality. Empirical validation across domains—such as text generation, music composition, and visual art—would demonstrate whether these metrics reliably predict when an agent’s contribution merits standalone authorial credit. Also, given the stochastic interplay of multi-agent pipelines, there is a need for algorithms that can disentangle and visualize each agent’s creative “fingerprint.” Described AI techniques could be adapted to highlight which components of an output were most influenced by which agent, thus attributing based on measurable statistical contributions.

\paragraph{Resource-Efficient Orchestration} While MAS promise remarkable creative capabilities through parallel specialization, they also introduce substantial computational overhead, making resource-efficient orchestration an urgent challenge \cite{creech2021resourceallocationdynamicmultiagent}. Naively spawning dozens of agents can lead to prohibitive latency, high cloud costs, and unsustainable energy consumption. \textbf{Self-Resource Allocation} \cite{amayuelas2025selfresourceallocationmultiagentllm} mechanism lets each agent budget its own compute, achieving near-optimal cost–performance trade-offs. \textbf{DynTaskMAS} \cite{yu2025dyntaskmasdynamictaskgraphdriven} leverages dynamic task graphs to asynchronously decompose workflows, reducing execution time by up to 33\% and improving utilization by 35\%. \textbf{MaAS}’s agentic supernet adapts architecture to each query, slashing inference costs to 6–45 \% of static systems \cite{zhang2025multiagentarchitecturesearchagentic}. 

Future research can explore adaptive agent pruning and distillation techniques that dynamically identify and deactivate or compress agents whose incremental contributions to a creative task fall below a meaningful threshold, yielding a leaner ensemble that retains quality while dramatically lowering computational overhead. Complementing this, meta-learning for orchestration policies could train a higher-order controller via meta-reinforcement learning to rapidly specialize scheduling strategies to new creative domains, such as narrative generation versus musical composition, using only a handful of trial interactions, thereby minimizing costly exploration in production. Finally, integrating human-in-the-loop orchestration channels will allow lightweight, real-time user feedback to signal when an intermediate creative draft meets subjective standards of “good enough,” enabling the system to halt or redirect further agent invocations and align resource consumption with human satisfaction rather than arbitrary performance metrics.

\paragraph{Longitudinal User Studies} In contrast to the abundance of controlled single-session evaluations, understanding how users engage with multi‐agent creative systems over extended periods remains a significant hurdle. Longitudinal investigations have revealed that users undergo an initial novelty phase before stabilizing their expectations and customizing AI workflows \cite{10.1145/3643834.3661587}. In educational settings, semester-long dialogues with ChatGPT demonstrated evolving revision strategies and satisfaction levels. This underscores that early positive impressions can change as learners develop mental models of AI partners \cite{han-etal-2024-recipe4u}. Temporal pattern analysis in collaborative writing revealed distinct AI reliance phases, where users gradually transition from exploratory interactions to purpose-driven selective assistance as trust and competence grow \cite{yang2024inkalgorithmexploringtemporal}.

Future work can focus on three directions. First, we need longitudinal studies that measure how users' creative abilities develop when collaborating with multi-agent systems. Through automated analysis and expert evaluation, these studies would track improvements in specific skills like narrative coherence or compositional technique. Second, researchers should investigate how users' and LLMs' understanding of different specialized agents (such as “plot architects” or “style editors”) evolves. This research would examine how these evolving perceptions affect which helpers humans or AI systems choose to collaborate with during different parts of the creative process. Third, systems that can customize agent teams for individual users should be developed: automatically introducing new agents, removing unhelpful ones, or adjusting existing agents based on the user's preferences and performance. This would create personalized creative partnerships that support each user's ongoing artistic development.

\section{Conclusion}
\label{sec:Conclusion} 

This survey examines the rise of LLM-based multi-agent systems for creative tasks. We propose a unified framework for collaborative workflows and analyze how agent proactivity influences idea generation. We then identify three key techniques that reliably enhance creative performance and review current evaluation methods. We also examine the persona’s impact on creativity, exploring how different persona profiles and granularity levels shape idea generation. Next, we survey the datasets used to measure creativity in multi-agent systems, categorizing them into psychological test datasets and task-specific collections. Finally, we discuss overarching challenges, such as adaptive initiative control, bias mitigation, scalable interaction protocols, and the lack of standardized benchmarks, and outline promising directions for future research. Our goal is to clarify this rapidly evolving field and support the development of transparent, effective systems that augment human creativity.


\section*{Limitations}
While this survey aims to provide a comprehensive overview of LLM-based creative multi-agent systems, several limitations remain that offer opportunities for future refinement. 

First, our focus on text and image modalities was intended to ensure depth of analysis, but it necessarily excludes other important interaction channels, such as audio~\cite{wu2024audiolanguagemodeling, kuan2024speechcopilotleveraginglargelanguage}, video~\cite{huang2024genmaccompositionaltexttovideogeneration}, and embodied robotics~\cite{10979714}, which may bring distinct challenges and opportunities for creative MAS.

Second, while we briefly discuss persona-related biases, we do not delve into broader ethical considerations. These include issues such as data licensing and provenance (e.g., the use of proprietary or copyrighted corpora), user privacy when agents log interactions or generate persistent memory traces, informed consent in human-agent data collection, and the environmental costs associated with large-scale multi-agent deployments.

Third, the majority of systems reviewed in this survey are developed and evaluated in English and rely heavily on Western-centric datasets. We do not cover how cultural norms, multilingual settings~\cite{10832317}, or low-resource languages affect agent design, creative expression, or evaluation standards. Addressing these dimensions is critical to building more inclusive, globally relevant systems that reflect diverse forms of creativity and collaboration.

\section*{Acknowledgements}

We gratefully acknowledge the assistance of \texttt{ChatGPT}, which was used during the writing process to refine phrasing, improve clarity, and ensure consistent tone throughout the paper. All content and ideas remain the responsibility of the authors, and no parts of the manuscript were autogenerated without human verification and editing.

\bibliography{custom}

\appendix
\label{sec:appendix}

\section{Proactivity Spectrum Supplementary \label{spectrum details}}

This section details how we classify the proactivity levels shown in Fig.~\ref{fig:fig2}. We classify agent proactivity along two dimensions—\emph{Process} and \emph{Decision Making}—using a rainbow scale from red (highest) to purple (lowest). We also note that the \emph{Planning} phase remains under-explored, likely due to LLM agents’ low confidence in autonomous planning.

The highest level of agent proactivity, marked in red, embodies a fully agent-driven pipeline. At this level, agents autonomously perform all tasks, including discussion, idea sharing, and peer review, without any human guidance or feedback. In \textbf{MaCTG} \cite{zhao2025mactg}, agents are assigned individual tasks and kick off the project on their own—they come up with ideas, write code, assemble components, validate results, and refine the output. From start to finish, the entire creative process runs without any human input. 

At the second-tier level of proactivity, represented in yellow and green, human intervention is slightly enhanced and helps shape the agent's output, resulting in more stable and predictable behavior \cite{10.1145/3613904.3642647, MellouliC24}.
\textbf{Co-Scientist} \cite{gottweis2025aicoscientist} enables human users to inject additional ideas into a shared workspace among agents, stimulating agents' divergent thinking during the creativity process. 
\textbf{CollabStory} \cite{collabstory} attempts to build a large-scale story creation database with minimal human effort. It provides LLM agents with brief human instructions and storylines from previous agents. These iterative human inputs have a latent influence on idea generation from agents, despite no direct outlines.

Human-Agent synergy, otherwise, leads to medium proactivity, characterized by peer-level collaboration. Both parties jointly engage in the \emph{Process} phase to enhance the diversity and feasibility of creative outputs. 
However, to prevent potential ethical hazards and unexpected outcomes, these frameworks tend to entrust the final evaluation to human users, thereby inevitably exhibiting low agent proactivity at \emph{Decision Making} \cite{10.1145/3613904.3642224, 10.1609/aiide.v18i1.21946}. 

In line with our definition of proactivity in Section ~\ref{sec:Agent Participation and Proactivity}, systems that follow data-driven processes or act beyond direct human instructions demonstrate a higher level of proactivity. In our classification, such cases are highlighted in blue, and this increase of proactivity is particularly prominent in the \emph{Process} phase. 
For example, \textbf{ContextCam} \cite{10.1145/3613904.3642129} not only receives iterative user requests during refinement but also incorporates environmental data collected from its sensors, such as weather conditions, camera input, and audio input. \textbf{Colin} \cite{ye2024colin} exhibits the agent proactivity through a different way. The system initiates the interaction questions to trace the understanding and idea of users, rather than relying on reactive prompt-based communication like typical Human-Agent synergy frameworks.

The purple-marked work shows relatively low proactivity in both phases. Humans mainly use the LLM agent to generate ideas from alternative perspectives, helping to fill in where human thinking might be limited. The system keeps its creative output strong by leaning on a solid human-driven backbone and manual evaluation. While the results are good, it also imposes an excessive load on designers and creators \cite{10.1145/3637361, 10.1145/3663384.3663398, lim2024rapid}.


\begin{table*}[htb]

  \small
  \centering
  \begin{tabular}{l l l l}
    \toprule
    \textbf{MAS Technique} & \textbf{Task Domain} & \textbf{Framework} \\
    \midrule
    \multirow{15}{*}{Divergent Exploration}
      & AUT and RAT                    & Long-Term Guidance                        \shortcite{Kumar2024HumanCI}\\
      & Character Design               & PersonaGym                                \shortcite{samuel2024personagymevaluatingpersonaagents}\\
      & Creative Writing               & Creativity Support (LLMs)
              \shortcite{chakrabarty2024creativitysupportagelarge}\\
      & Humor Co-Creation              & Meme Alone
                       \shortcite{Wu_2025}\\
      & Idea Generation                & Co-GPT Ideation                           \shortcite{lim2024rapid}\\
      & Idea Generation                & Group-AI Brainwriting                     \shortcite{10.1145/3613904.3642414}\\
      & Idea Generation                & Virtual Canvas                            \shortcite{10.1145/3663384.3663398}\\
      & Image Generation          & ContextCam                                \shortcite{10.1145/3613904.3642129}\\
      & Interior Color Design Ideation          & C2Ideas                                   \shortcite{10.1145/3613904.3642224}\\
      & Research Ideation              & Ideation Co-Pilot
                       \shortcite{nigam-etal-2024-interactive}\\
      & Research Ideation              & PersonaFlow                               \shortcite{liu2024personaflowboostingresearchideation}\\
      & Scientific Research Co-creation& VirSci                                    \shortcite{su2025headsbetteroneimproved}\\
      & Sketches Generation            & StoryDrawer                               \shortcite{10.1145/3491102.3501914}\\
      & Story Generation               & ICCRI                                     \shortcite{10.5555/3721488.3721531}\\
      & Story Generation               & SPARKIT                                   \shortcite{10.1007/978-3-031-71152-7_1}\\
    \midrule

    \multirow{10}{*}{Iterative Refinement}
      & Character Design               & CharacterMeet                             \shortcite{10.1145/3613904.3642105}\\
      & Debating (Fairness)            & Multi-Agent Debate                        \shortcite{liang-etal-2024-encouraging}\\
      & Hallucination Mitigation       & Hallucination Mitigation
      \shortcite{lin2024interpretingmitigatinghallucinationmllms}\\
      & Legal Consultation             & LawLuo                                    \shortcite{sun2024lawluomultiagentcollaborativeframework}\\
      & LLM Pipeline Generation        & ChainBuddy                                \shortcite{Zhang_2025}\\
      & Product Design                 & DesignGPT                                 \shortcite{10494260}\\
      & Scientific Research Co-creation& Baby-AIGS-MLer                            \shortcite{anonymous2024aml}\\
      & Screenwriting                  & HoLLMwood                                 \shortcite{DBLP:journals/corr/abs-2406-11683}\\
      & Sketches Generation            & CICADA                                    \shortcite{10179136}\\
      & Social Simulation              & Generative Agents                         \shortcite{10.1145/3586183.3606763}\\
    \midrule

    \multirow{22}{*}{Collaborative Synthesis}
      & Agent Benchmarking             & TheAgentCompany                           \shortcite{xu2024theagentcompanybenchmarkingllmagents}\\
      & Agent Collab. Visualisation    & AgentCoord                                \shortcite{pan2024agentcoordvisuallyexploringcoordination}\\
      & Cognitive Synergy             & Solo Performance Prompting        \shortcite{wang2024unleashingemergentcognitivesynergy}\\
      & Creative Writing               & Human-AI Co-creativity                    \shortcite{10.1145/3637361}\\
      & Creativity Simulation          & Creative Agents                           \shortcite{imasato2024creativeagentssimulatingsystems}\\
      & Formal Proof                   & ProofNet                                  \shortcite{azerbayev2023proofnet}\\
      & Program Design                 & Beyond Code Generation                    \shortcite{Zamfirescu_Pereira_2025}\\
      & Recommendation                 & Agent4Rec             \shortcite{10.1145/3626772.3657844}\\
      & Research Ideation              & CoQuest                                   \shortcite{10.1145/3613904.3642698}\\
      & Research Peer Review           & MARG                                      \shortcite{darcy2024margmultiagentreviewgeneration}\\
      & Scientific Peer Review         & AgentReview                               \shortcite{jin-etal-2024-agentreview}\\
      & Scientific Research Co-creation& CrewAI                                    \shortcite{venkadesh2024unlocking}\\
      & Scientific Research Co-creation& Co-Scientist                              \shortcite{gottweis2025aicoscientist}\\
      & Silhouette Generation          & Human-Machine Co-Creation                 \shortcite{lataifeh2024human}\\
      & Software Engineering           & ChatDev                                   \shortcite{qian-etal-2024-chatdev}\\
      & Software Engineering           & MaCTG                                     \shortcite{zhao2025mactg}\\
      & Story Generation               & CollabStory                               \shortcite{collabstory}\\
      & Story Generation               & Colin                                     \shortcite{ye2024colin}\\
      & Story Generation               & Mathemyths                                \shortcite{10.1145/3613904.3642647}\\
      & Story Generation               & StoReys                                   \shortcite{MellouliC24}\\
      & Story Generation               & StoryVerse                                \shortcite{10.1145/3649921.3656987}\\
      & UI Prototyping                & MAxPrototyper                             \shortcite{yuan2024maxprototypermultiagentgenerationinteractive}\\
    \bottomrule
  \end{tabular}

  \caption{Overview of representative MAS frameworks categorized by their core creative techniques—Divergent Exploration, Iterative Refinement, and Collaborative Synthesis.}
  \label{tab:mas-techniques-extended}
\end{table*}
\begin{table*}[htb]
    \centering\small
    \begin{tabular}{
    p{4cm}p{6cm}p{5cm}
    }
    \toprule
    \textbf{Task} & \textbf{Task Description} & \textbf{Available Datasets} \\
    \midrule
    Problem-Solving in Physically Grounded Scenarios & Test multi-agent's ability to think resourcefully and act creatively in novel physical situations. & MacGyver \cite{tian2024macgyver} \\
    \hline
    Creative Writing & Evaluate the writing skills and collaborative abilities of multi-agents.   & Human-AI co-writing Stories \cite{chakrabarty2024creativitysupportagelarge}, CollabStory \cite{collabstory} \\
    \hline
    Music Genre & Evaluate the multi-agents in robot dance creation. &\cite{de2024large} \\
    \hline
    Character Design & Evaluating the creativity of multi-agent systems in visualizing and generating new characters. & \cite{lataifeh2024human} \\
    \hline
    QA problem & Open-ended question task. &TriviaQA \cite{joshi2017triviaqa}, QA in Game simulation \cite{10.1145/3586183.3606763}, GPQA Diamond Set \cite{rein2024gpqa} \\
    \hline
    Codenames Task & Evaluate the models' ability to identify words associated with a given word. & \cite{srivastava2022beyond} \\
    \hline
    Mathematical Formal Proof Generation and Verification & Test the model's ability of  autoformalization and formal proving
of undergraduate-level mathematics. &\cite{azerbayev2023proofnet} \\
    \hline
    Idea Generation  & Quantitatively evaluate
and compare the ideas generated by LLMs. &AI Idea Bench 2025\cite{qiu2025aiideabench2025}, AMiner Computer Science Dataset \cite{tang2008arnetminer}, LiveIdeaBench \cite{ruan2024liveideabench} \\
    \hline
    Fairness-Aware Debating & Evaluate the ethical and practical implications of automated decision-making systems in the justice system. & COMPAS dataset \cite{propublica2016compas} \\
    \bottomrule
    \end{tabular}
    \caption{Creative tasks along with their associated datasets. Tasks that lack datasets and rely primarily on user studies are not included.}
    \label{tab:datasets}
\end{table*}
\begin{table*}[htb]
    \centering
    \small
    \begin{tabular}{lcccl}
    \toprule
     \bf Framework & \bf Granularity & \bf Method & \bf Persona Example\\
    \midrule
         Solo Performance Prompting~\shortcite{wang2024unleashingemergentcognitivesynergy}
        &  \multirow{8}{*}{ Coarse} &  Model-Generated 
        &  Self-Defined  &   \\
         LLM Discussion~\shortcite{lu2024llmdiscussionenhancingcreativity}&  &  Model-Generated &  Self-Defined 
        \\
         PersonaGym~\shortcite{samuel2024personagymevaluatingpersonaagents} &      &  Human-Defined &  Self-Defined  \\
         Baby-AIGS-MLer~\shortcite{anonymous2024aml} &  &  Human-Defined &  Assistant  \\
         SPARKIT~\shortcite{10.1007/978-3-031-71152-7_1} &  &  Human-Defined &
         Self-Defined \\
         Multi-Agent Debate~\shortcite{liang-etal-2024-encouraging}& &  Human-Defined &  Debater &  \\
         Acceleron~\shortcite{nigam-etal-2024-interactive} & &  Human-Defined &  Mentor \& Colleague &   \\
         ChainBuddy~\shortcite{Zhang_2025} & &  Human-Defined &  Mentor \& Planner &  \\

    \midrule
            TheAgentCompany~\shortcite{xu2024theagentcompanybenchmarkingllmagents} &   \multirow{9}{*}{ Medium-Coarse}&  Model-Generated &  Company Employee \\
             HoLLMwood~\shortcite{DBLP:journals/corr/abs-2406-11683}&  &  Human-Defined
            &  Artist &  \\
             TRIZ Agents~\shortcite{szczepanik2025triz}&   &  Human-Defined &  Problem Solver \\
             Co-Scientist~\shortcite{gottweis2025aicoscientist} &   &  Human-Defined &  Researcher \\
             MaCTG~\shortcite{zhao2025mactg}&  &  Human-Defined &  Programmer  \\
             DesignGPT~\shortcite{10494260}&  &  Human-Defined &  Self-Defined &  \\
             CoQuest~\shortcite{10.1145/3613904.3642698} &   &  Human-Defined &  Researcher &  \\
             LawLuo~\shortcite{sun2024lawluomultiagentcollaborativeframework} & &  Human-Defined &  Lawyer  \\
             MARG~\shortcite{darcy2024margmultiagentreviewgeneration} & &  Human-Defined &  Expert  \\
            
    \midrule
        
             PersonaFlow~\shortcite{liu2024personaflowboostingresearchideation}&  \multirow{9}{*}{ Fine} &  Data-Derived & Researcher \\
             VirSci~\shortcite{su2025headsbetteroneimproved}&  &  Data-Derived &  Researcher \\
             Agent4Rec~\shortcite{10.1145/3626772.3657844} & 
            &  Data-Derived &  Media User \\
             PersonaLLM~\shortcite{jiang2024personallminvestigatingabilitylarge} &  &  Model-Generated &  Self-Defined \\
             The Power of Personality~\shortcite{duan2025powerpersonalityhumansimulation} &  &  Model-Generated &  Self-Defined \\
             Creative Agents~\shortcite{imasato2024creativeagentssimulatingsystems} &  
            &  Model-Generated &  Artist \\
             CoAGent~\shortcite{zheng2023synergizinghumanaiagencyguide}  &  &  Model-Generated &  Self-Defined \\
             Generative Agents~\shortcite{10.1145/3586183.3606763} & &  Human-defined &  Sandbox Character  \\
             AgentReview~\shortcite{jin-etal-2024-agentreview} & &  Human-defined &  Reviewer \\
    \bottomrule
    \end{tabular}
    \caption{ Summary of agent profile granularity and generation methods in MAS, with each paradigm’s role definition and paper citation. ‘Self-defined’ personas grant agents the freedom to adopt diverse characters, promoting flexible collaboration and creative innovation.}
    \label{tab:persona_category}
\end{table*}
\begin{table*}[htb]
  \small
    \begin{tabular}{llp{3.5cm}p{3.5cm}}
    \toprule
     \textbf{Paper} & \textbf{Task}  & \textbf{Subjective}& \textbf{Objective}\\
    \midrule 
     \citet{Kumar2024HumanCI} &  AUT and RAT &  TTCT, Boden's Criterion, and others &  Semantic similarity\\
     \citet{duan2025powerpersonalityhumansimulation} &  AUT and others &  TTCT &  - \\
     \citet{lu2024llmdiscussionenhancingcreativity} &  AUT and others &  TTCT &  - \\
     \citet{10747324} &  Conceptual Design &  TTCT &  - \\
     \citet{lim2024rapid} &  Idea Generation &  TTCT &  - \\
     \citet{10.1145/3613904.3642414} &  Idea Generation &  Innovation, Insightfullness, and others &  Semantic Similarity \\
     \citet{sun2024lawluomultiagentcollaborativeframework} &  Legal Consultation &  Personalization and Professionalism &  - \\
     \citet{azerbayev2023proofnet} &  Mathematical Proving &  - &  BLEU Score \\
     \citet{10494260} &  Product Design &  Novelty, Completeness, and Feasibility &  - \\
     \citet{wan2025usinggenerativeaipersonas} &  Plot Generation &  - &  Semantic Similarity \\
     \citet{chakrabarty2024creativitysupportagelarge} &  Poem Writing &  Fluency and Creativity &  - \\
     \citet{darcy2024margmultiagentreviewgeneration} &  Paper Review Generation &  Specificity and Overall Rating &  - \\
     \citet{10.1145/3613904.3642698} &  Research Ideation &  Boden's criterion &  - \\
     \citet{liu2024personaflowboostingresearchideation} &  Research Ideation &  Creativity, Usefulness, and Helpfulness &  - \\
     \citet{DBLP:journals/corr/abs-2406-11683} &  Screenwriting &  Interestingness, Relevance and others &  Entropy-n, Self-BLEU and others \\
     \citet{gottweis2025aicoscientist} &  Scientific Research Co-creation &  Novelty and Impact &  - \\
     \citet{10.1145/3635636.3664627} &  Scientific Research Co-creation &  Novelty, Specificity, and others &  Semantic Similarity \\
     \citet{su2025headsbetteroneimproved} &  Scientific Research Co-creation &  Novelty, Clarity, Feasibility &  Semantic Euclidean Distance\\
     \citet{10.1145/3613904.3642647} &  Story Generation &  Readability, Perceived Creativity, and others &  - \\
     \citet{MellouliC24} &  Story Generation &  Interactivity, Coherence, and others &  Self-BLEU \\
     \citet{10.1609/aiide.v18i1.21946} &  Story Generation &  Goal completion and Satisfication &  - \\
     \citet{collabstory} &  Story Generation &  Creativity &  Entropy and others\\
     \citet{10.5555/3721488.3721531} &  Story Generation &  TTCT &  - \\

    \midrule
         \citet{10.1145/3613904.3642224} &  Interior Color Design Ideation &  Inspiring, Reasonableness, and others &  - \\
         \citet{venkatesh2025creacollaborativemultiagentframework} &  Image Editing \& Generation &  Expressiveness, Aesthetic appeal, and others &  CLIP scores and others\\
         \citet{Wu_2025} &  Meme Generation &  Funniness, Creativity, and Shareability &  - \\
         \citet{10.1145/3491102.3501914} &  Sketches Generation  &  TTCT &  - \\
         \citet{10179136} &  Sketches Generation &  TTCT &  FID, TIE ,and Semantic loss \\
         \citet{10.1145/3581641.3584095} &  Sketches Generation &  Novelty and Surprise within MICSI &  - \\
         \citet{lataifeh2024human} &  Silhouette Generation &  Designer's review &  FID \\
         \citet{yuan2024maxprototypermultiagentgenerationinteractive} &  UI Prototype Generation &  - &  FID and Generation Diversity \\

    \bottomrule
    \end{tabular}
    \caption{Output evaluation methods employed across various tasks. The upper section details evaluations for text generation tasks, while the lower section focuses on image generation tasks.}
    \label{tab:evaluation}
\end{table*}

\end{document}